\documentclass[useAMS,usenatbib,usegraphicx]{mn2e}

\newcommand{\dd}{{\rm d}}
\newcommand{\deriv}[2]{\frac{\dd#1}{\dd#2}}

\newcommand{\code}[1]{{\tt #1}}

\newcommand{\Revised}[1]{{#1}}

\title[Self-gravitating photoionized gas]
{Photoionized gas in hydrostatic equilibrium: the role of gravity}

\author[Y.~Ascasibar and A.~D\'{i}az]
{
  Y.~Ascasibar\thanks{E-mail: yago.ascasibar@uam.es} and A.~I.~D\'{i}az\\
  Departamento de F\'{i}sica Te\'{o}rica, Universidad Aut\'{o}noma de Madrid, 28049 Madrid, Spain
}

\date{Draft version 2.0 (\today)}

\begin{document}

\maketitle

\begin{abstract}
We present a method to include the effects of gravity in the plasma physics code {\sc Cloudy}.
More precisely, a term is added to the desired gas pressure in order to enforce hydrostatic equilibrium, accounting for both the self-gravity of the gas and the presence of an optional external potential.
As a test case, a plane-parallel model of the vertical structure of the Milky Way disk near the solar neighbourhood is considered.
It is shown that the gravitational force determines the scale height of the disk, and it plays a critical role in setting its overall chemical composition.
However, other variables, such as the \Revised{shape of incident continuum and the intensity of the} Galactic magnetic field, strongly affect the \Revised{predicted} structure.
\end{abstract}

\begin{keywords}
ISM:structure -- methods:numerical -- \Revised{gravitation -- radiative transfer}
\end{keywords}

\section{Introduction}
\label{secIntro}

Many astrophysical objects, from the scale of individual stars to the hot plasma of the intracluster medium, can be approximately described as being in hydrostatic equilibrium with the underlying gravitational potential.
The density and temperature of the gas are related to the enclosed mass by the condition that the net acceleration vanishes.
Thus, the hydrostatic equilibrium equation reads
\begin{equation}
 \frac{1}{\rho(x)}\deriv{P(x)}{x} = g(x)
\label{eqHydro_usual}
\end{equation}
where $P$ denotes the total (gas, magnetic, turbulent, cosmic-ray, and radiation) pressure, and $x$ is the relevant spatial coordinate.
In spherical coordinates, $x \equiv r$, and the gravitational acceleration
\begin{equation}
 g(r) = - \frac{GM(r)}{r^2} = - \frac{4\pi G}{r^2} \int_0^r x^2 \rho(x)\ \dd x
\label{eqGrav_sphere}
\end{equation}
is determined by the total mass $M(r)$ contained within radius $r$.
For a symmetric plane-parallel atmosphere, the relevant coordinate is the distance to the mid plane, $z$, and the gravitational acceleration is
\begin{equation}
 g(z) = - 2\pi G\,\Sigma(z) = - 4\pi G \int_0^z \rho(x)\ \dd x
\label{eqGrav_plane}
\end{equation}
In either case, the solution to equation~(\ref{eqHydro_usual}) is
\begin{equation}
 P(x) = P_0 + \Delta P_{\rm grav}(x) \equiv P_0 + \int_0^x \rho(u)\ g(u)\ \dd u
\label{eqHydro}
\end{equation}
where $P_0$ is the central pressure.
There are many situations of astrophysical interest where $\Delta P_{\rm grav}(x) \ll P_0$, and the hydrostatic equilibrium condition results in an almost isobaric atmosphere.
However, depending on the total mass and the characteristic temperature of the gas, the gravitational field may be strong enough to have a significant effect on the structure of the system.

One example of a photoionized gas where the gravitational acceleration is not negligible is the Galactic interstellar medium (ISM).
Although interstellar matter only accounts for a relatively small fraction of the total disk mass, it plays a crucial role on the formation and evolution of the Galaxy.
New stars are formed from the cold molecular phase of the ISM, and the energy input from the stellar population \Revised{influences} the structure and composition of the interstellar gas.
The complex interplay between gas, stars, cosmic rays, and magnetic fields is responsible for the rich structure, dynamics, and observational phenomenology of our Galaxy \citep[see e.g.][]{Ferriere01}.

The present work describes the implementation of gravitational hydrostatic equilibrium in the plasma physics code {\sc Cloudy}, last described by \citet{Ferland+98}.
Numerical details are given in Section~\ref{secCloudy} and Appendix~\ref{secUserGuide}.
Section~\ref{secResults} discusses the results obtained for the vertical structure of the Milky Way disk in the solar neighbourhood, and a brief summary is given in Section~\ref{secConclusions}.

\section{Numerical implementation}
\label{secCloudy}

The {\sc Cloudy} photoionization code computes the ionization structure, level populations, electron temperature, and observable spectrum of a gas, given its chemical composition and the spectrum of an ionizing source.
The program allows the user to set the gas density at the illuminated face and to choose from several prescriptions to specify how it varies with distance.

One of these prescriptions is \code{constant pressure}, which may refer to the gas or the total pressure
\begin{equation}
P(x) = P_{\rm gas} + P_{\rm ram} + P_{\rm turb} + P_{\rm mag} + P_{\rm lines} + \Delta P_{\rm rad}
\end{equation}
where $P_{\rm gas}=nkT$ is the thermal pressure, $P_{\rm ram}=\rho v_{\rm wind}^2$ and $P_{\rm turb}=\rho v_{\rm turb}^2/2$ arise from uniform and turbulent motion, respectively, $P_{\rm mag}=\frac{B^2}{8\pi}$ is the magnetic pressure, $P_{\rm lines}$ is the radiation pressure of trapped emission lines \citep{FerlandElitzur84,ElitzurFerland86} and $\Delta P_{\rm rad} \equiv \int \rho\,a_{\rm rad}\,\dd x$ compensates for the acceleration $a_{\rm rad}$ due to the absorption of the incident radiation.
This term is analogous to $\Delta P_{\rm grav}$ and, to some extent, one could argue that the command should already be named ``hydrostatic'' rather than ``constant pressure'' even if it did not account for the effect of gravity.
Since $a_{\rm rad}$ is positive (the momentum of the absorbed photons points outwards), $\Delta P_{\rm rad}$ induces a positive pressure gradient (see Figure~\ref{figConstantP}), whereas gravity must be balanced by a negative gradient.

We have modified the \code{constant pressure} option in version 08.00 of {\sc Cloudy} to account for the gravitational term $\Delta P_{\rm grav}$.
Source code is available upon request, and a brief user guide is provided in Appendix~\ref{secUserGuide}.
The required modifications are relatively minor, and they are not expected to incur in any side effect on the functionality of the algorithm.
\Revised{They are currently being implemented in the development version of the code and will be available in the next official release\footnote{The current stable version of {\sc Cloudy} can be downloaded from http://wiki.nublado.org/wiki/Download}.}

First, new variables are created to store the values of $\Delta P_{\rm grav}$ (initialized to 0 at the illuminated face), the symmetry of the problem (spherical or plane-parallel) and the mass (0 by default) of an additional component located at the centre.

As {\sc Cloudy} moves away from the illuminated face, $\Delta P_{\rm grav}$ is updated according to
\begin{equation}
 \Delta P_{i+1} = \Delta P_i + \rho_i\, g_i\, h_i
\end{equation}
where $\Delta P_i$ and $\Delta P_{i+1}$ denote the old and new values, while $\rho_i$, $g_i$, and $h_i$ are the density, gravitational acceleration, and thickness of the last zone, respectively.
The gravitational acceleration is computed according to expression~(\ref{eqGrav_sphere}) or (\ref{eqGrav_plane}) depending on the geometry.
{\sc Cloudy} keeps track of the cumulative mass $M(r)$ and the gas column density $\Sigma(z)$ in each case.
Additional mass components or any external (gravitational or not) potential can be trivially accounted for, simply adding the appropriate contribution to the local acceleration $g_i$.

Finally, one has to modify the pressure loop so that the the term $\Delta P_{\rm grav}$ is added right after $\Delta P_{\rm rad}$ to balance the gravitational force.
Then, the code proceeds exactly as in the original constant pressure case, trying different values of the gas density until the deviation from the desired pressure is smaller than the requested tolerance.

\section{The Milky Way disk}
\label{secResults}

In order to test the proposed implementation of hydrostatic equilibrium, we consider a simple model of the ISM in the solar neighbourhood, assuming a plane-parallel geometry that is symmetric with respect to the Galactic mid plane, $z=0$.
The gas chemical composition \citep{CowieSongaila86,SavageSembach96,Meyer+98}, including dust grains \citep{Mathis+77}, is set by the \code{abundances ISM} command, while cosmic ray heating and ionization are treated following \citet{FerlandMushotzky84} with the instruction \code{cosmic ray background} (see Hazy\footnote{Available at http://www.nublado.org}, the {\sc Cloudy} documentation, for details).

\begin{figure}
\centerline{ \includegraphics[width=0.5\textwidth]{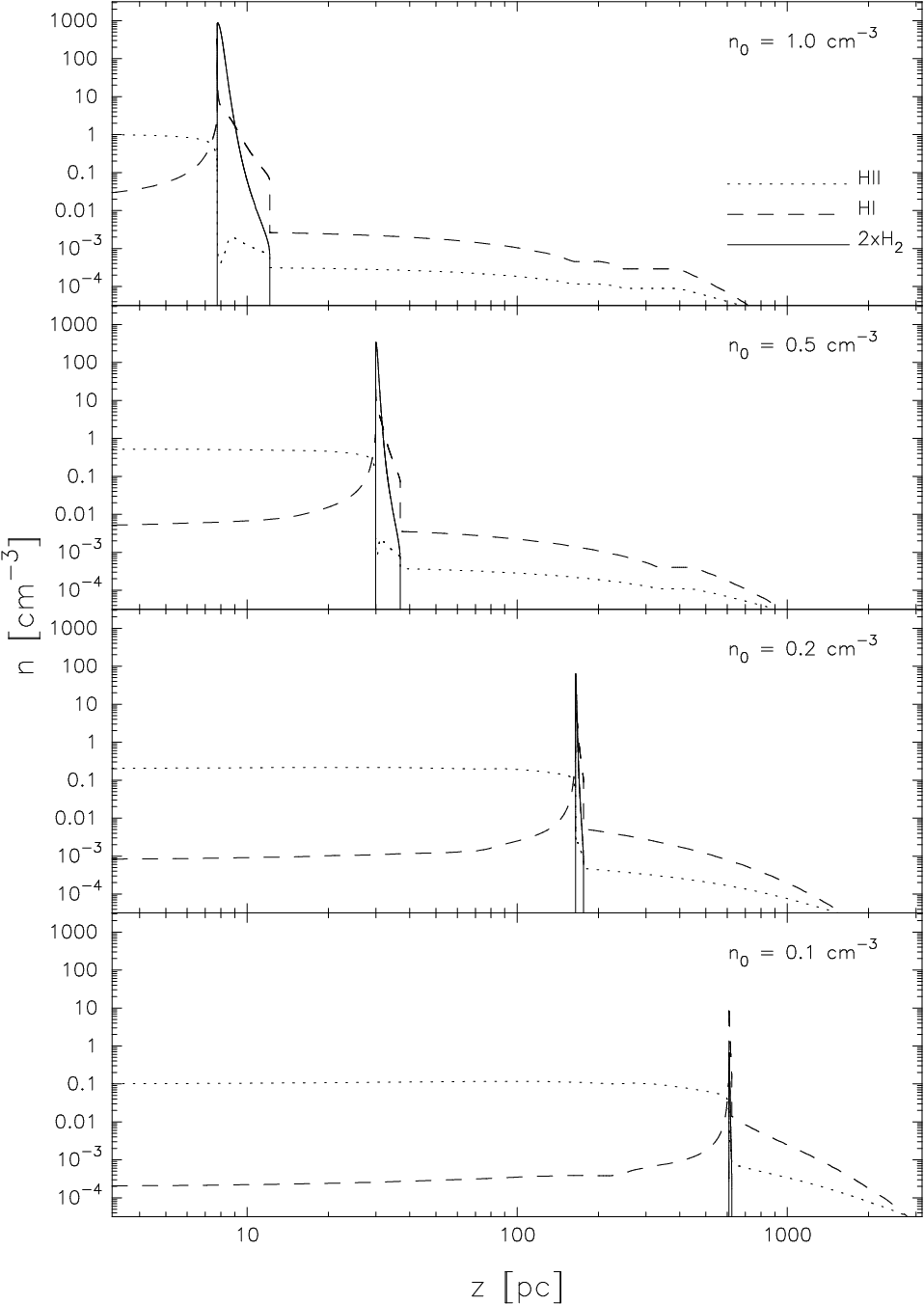} }
\caption
{
Molecular, neutral, and ionized Hydrogen density as a function of height $z$ with respect to the disk mid plane (dotted, solid, and dashed lines, respectively) for different values of the mid-plane density $n_0$.
}
\label{figN}
\end{figure}

\begin{figure}
\centerline{ \includegraphics[width=0.5\textwidth]{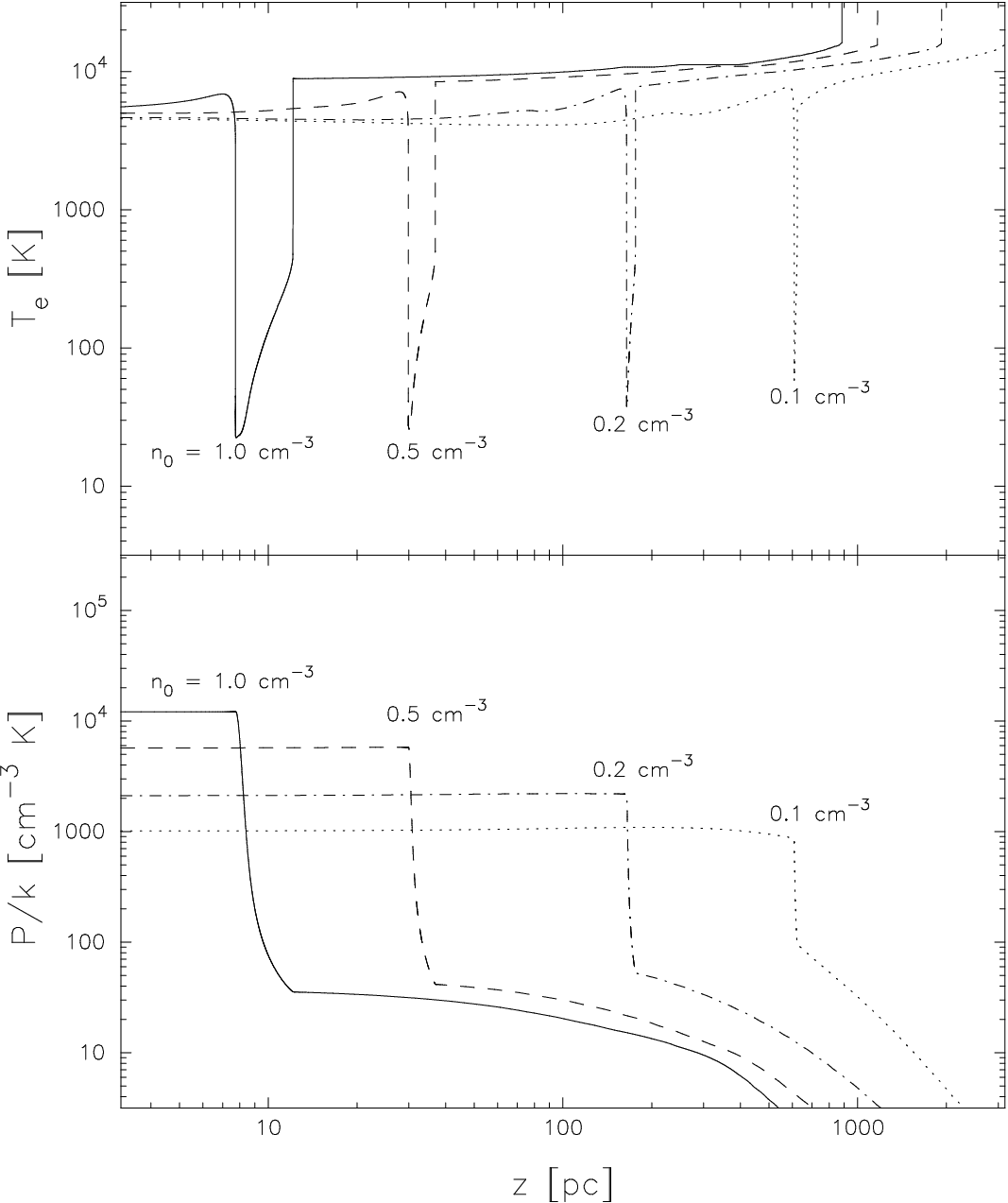} }
\caption
{
Gas temperature (top) and pressure (bottom) along the $z$-axis for values of the mid-plane density $n_0=1.0$, $0.5$, $0.2$, and $0.1$~cm$^{-3}$.
}
\label{figP}
\end{figure}

Given a mid-plane density $n_0$, the vertical structure of the disk is self-consistently computed by the program under the assumptions of ionization, thermal, and hydrostatic equilibrium.
An important limitation is that the gas is described by a single density and temperature that may vary with distance, but the coexistence of different phases at the same point is not implemented \citep[see e.g.][for examples of multiphase models of galaxy disks in hydrostatic equilibrium]{BoularesCox90,SilichTenorio-Tagle98,Ferriere01,Cox05}.
Another simplification is that the radiation field is assumed to strike the system from one side (in our case, the mid plane), whereas in reality the stars are distributed across a scale height comparable to that of the gaseous component \citep[e.g.][]{MillerCox93,Robin+03}.
With these caveats in mind, it is not surprising that our simple model does not provide a perfect match to the available observational data.
It should be able, though, to give a rough estimate of the \emph{volume-averaged} densities and scale heights of the different ISM phases.
More importantly, this kind of models offers some quantitative insight on the role played by the different physical mechanisms that may be at stake.
Although the present study is mostly concerned with gravity, we will also explore the influence of the spectral energy distribution of the incident radiation, as well as the presence of an external potential and a magnetic field.

\subsection{The role of gravity}

To illustrate the effect of gravitational acceleration on the equilibrium structure of a generic photoionized region, we ran a first model where the incident continuum consists of the cosmic microwave background and the interstellar radiation field, interpolated from the measurements of \citet{Black87} by means of the \code{table ISM} command.

\begin{figure}
\centerline{ \includegraphics[width=0.5\textwidth]{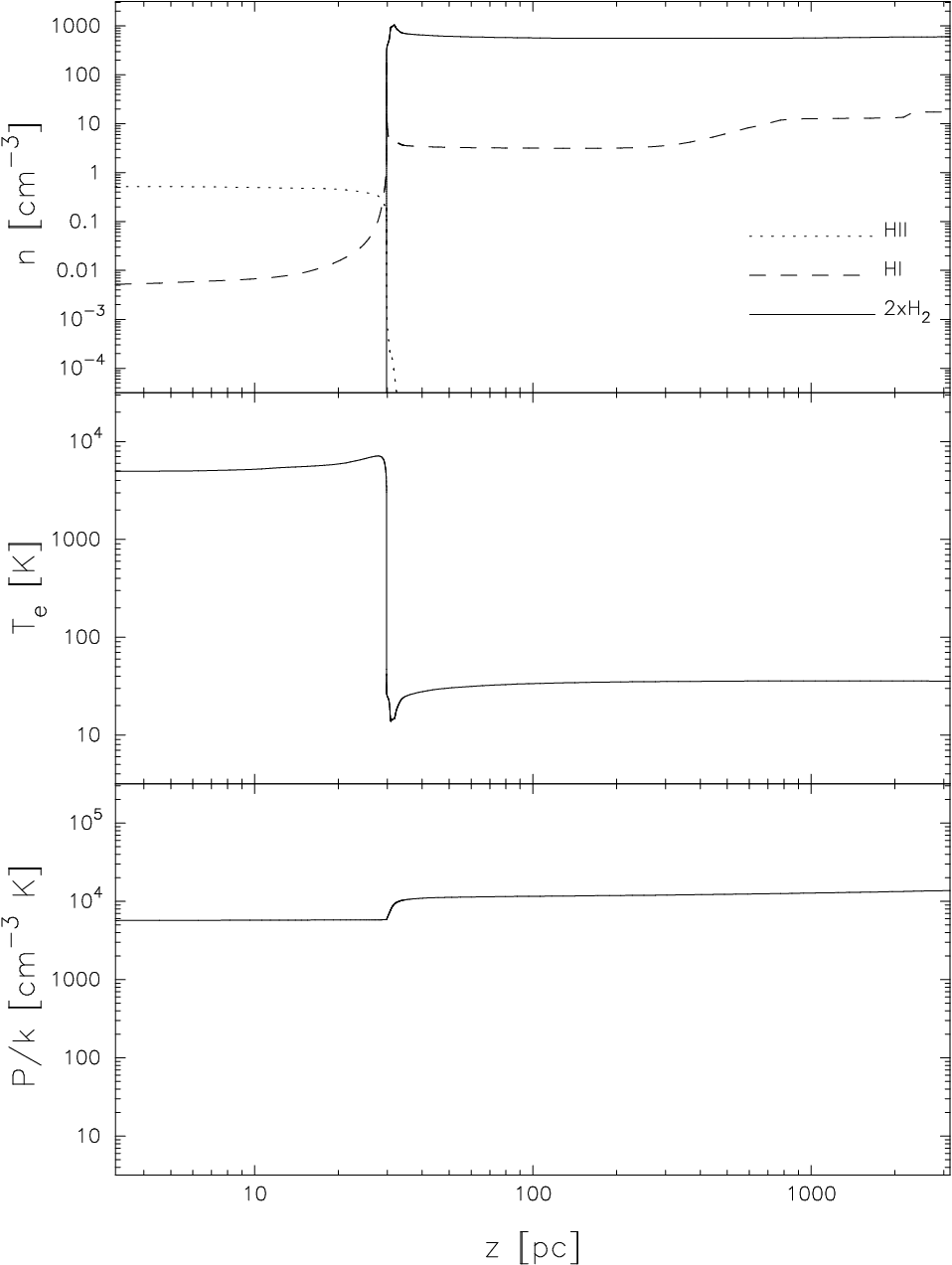} }
\caption
{
Density of the different Hydrogen phases (top), electron temperature (middle) and total pressure (bottom) in the standard \code{constant pressure} case.
}
\label{figConstantP}
\end{figure}

Results for different values of the central density $n_0$ are plotted in Figure~\ref{figN}.
In all cases, there is an inner region where the gas is almost fully ionized, a thin layer of molecular and atomic Hydrogen at high densities, and an outer region of low-density gas where Hydrogen is predominantly neutral, becoming more and more ionized at large distances as density decreases.
Gas temperature and pressure as a function of $z$ are shown in Figure~\ref{figP}.
The temperature of the cold dense phase is of the order of $10-100$~K, while the diffuse gas ranges from $5000$ to $10000$~K.
Gas pressure is roughly constant up to the ionization front, but it decreases rapidly through the molecular layer due to the high gas densities (above $100$~cm$^{-3}$).
The pressure stabilizes again in the outer region, although the gravitational acceleration imposes a negative pressure (and density) gradient.
The effect increases with distance; at some point, the cumulative gas mass and therefore the gravitational force are so high that both density and pressure decay extremely fast.
Gas temperature, on the other hand, increases steadily towards the Galactic halo.

This structure does not provide a realistic description of the local ISM.
On the one hand, the approximation that all the radiation comes from an infinitely thin disk at the mid plane creates a photoionized region at low $z$.
Although there is indeed a thin layer of ionized gas near the Galactic plane, composed of both discrete {\sc Hii} regions and diffuse gas \citep[see e.g.][]{Paladini+05}, its contribution near the solar neighbourhood is negligible \citep[e.g.][]{TaylorCordes93,CordesLazio_02}.
On the other hand, neither the position nor the thickness of the neutral phases agree with observations of the Milky Way ISM \citep{BoularesCox90,Ferriere01,NakanishiSofue03,NakanishiSofue06}.
Nevertheless, a central ionized region may be present in several astrophysical scenarios\footnote{Note that such an equilibrium configuration is, in any case, Rayleigh-Taylor unstable. Cold gas blobs and filamentary structures would slowly detach from the dense, neutral material and fall towards the low-density, ionized region, whereas hot gas bubbles would tend to rise buoyantly through the neutral phase.}, and therefore it is interesting to test the influence of gravity in this particular example.

The role of gravity can be assessed by comparison with the standard \code{constant pressure} solution, depicted in Figure~\ref{figConstantP} for a mid-plane density $n_0=0.5$~cm$^{-3}$.
The vertical structure of the disk is basically the same (i.e. gravity does not play a significant role) up to the ionization front.
There, the temperature decreases sharply, and the density has to rise accordingly to compensate.
In the standard implementation, pressure increases in order to balance the radiative acceleration ($\Delta P_{\rm rad}>0$).
The effect is very significant near the ionization front, but it rapidly decreases as photons become absorbed.
Although the pressure keeps increasing monotonically with $z$, the slope is quite shallow, and the physical state of the gas in the outer region (density, chemical composition, and temperature) remains approximately constant.
Thus, the main effect of gravity on a plane-parallel atmosphere is to set the extent of the molecular layer and the total scale height of the system.
Without gravity, the neutral (mostly molecular) layer would simply extend to infinity, as can be readily seen in Figure~\ref{figConstantP}.

\subsection{The incident radiation field}

\begin{figure}
\centerline{ \includegraphics[width=0.5\textwidth]{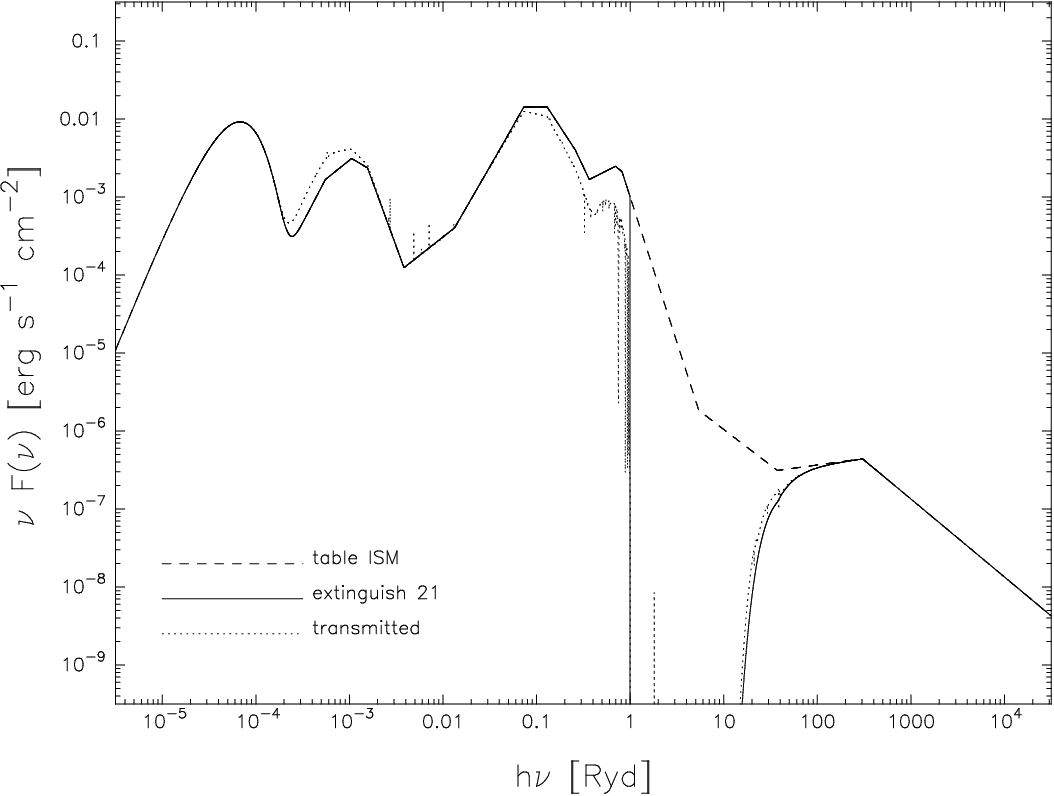} }
\caption
{
Incident continuum given by the \code{table ISM} command (dashed line), absorption of ionizing photons by the \code{extinguish} command with a Hydrogen column density of $10^{21}$~cm$^{-2}$ (solid line), and propagation through a uniform gas slab with constant density of $1$~cm$^{-3}$ and the same Hydrogen column density (dotted line).
}
\label{figContinua}
\end{figure}
\begin{figure}
\centerline{ \includegraphics[width=0.5\textwidth]{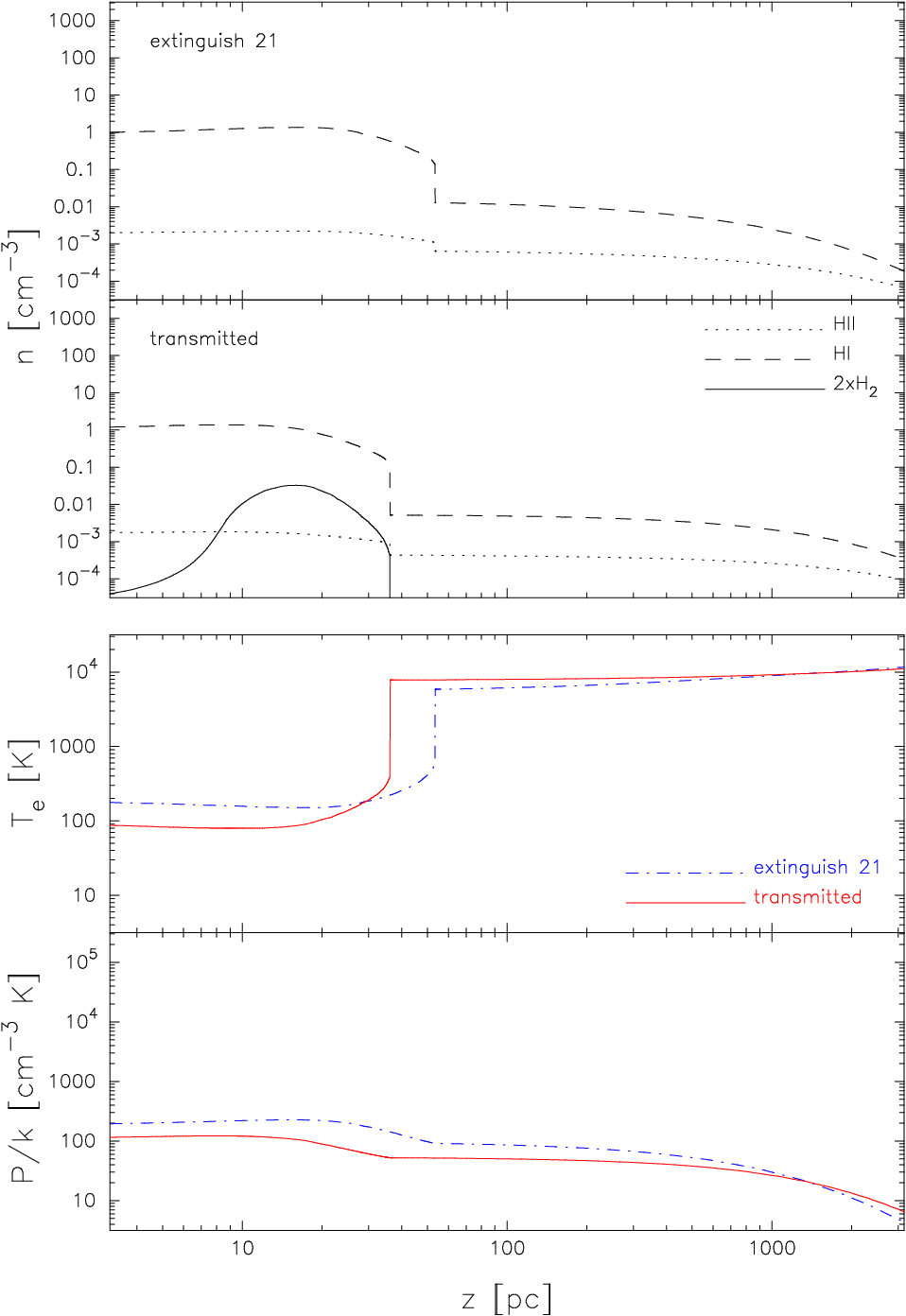} }
\caption
{
Effects of the incident continuum on the vertical structure of a model with $n_0=1$~cm$^{-3}$.
The \code{table ISM} case is depicted in Figures~\ref{figN} and~\ref{figP}.
The two top panels show the molecular, neutral and ionized Hydrogen densities for the incident radiation fields obtained by applying the \code{extinguish} command and the transmission through a uniform slab.
The third and fourth panels show the gas temperature and the total pressure in each case.
}
\label{figExtTrans}
\end{figure}

As discussed in the documentation of the \code{table ISM} command, ionizing radiation between 1 and 4 Rydberg is heavily absorbed by the neutral Hydrogen in the ISM so that few such photons exist, at least in the Galactic plane.
In order to describe the actual radiation field incident on a typical region in the Galactic plane, the use of the \code{extinguish} command is recommended.
However, it is also stated that this command should not be used except as a quick test, and a more physical extinction of the continuum may be accomplished by transmitting it through a model of the absorbing slab.

For the sake of simplicity, we assume a uniform density of $1$~cm$^{-3}$ and a total column density of $10^{21}$~cm$^{-2}$, appropriate for the Milky Way disk \citep{DickeyLockman90}.
The main difference between the resulting continua, shown in Figure~\ref{figContinua}, is that the \code{extinguish} command does not affect the energy band between 0.1 and 1 Rydberg, responsible for molecular Hydrogen photodissociation.
When the incident radiation field is propagated through the ISM, many of these ultraviolet photons become absorbed by the dust component and re-emitted in the far infrared.

The details of the interstellar radiation field have a dramatic impact on the vertical structure computed by the code.
Figure~\ref{figExtTrans} compares the results obtained for $n_0=1$~cm$^{-3}$ using the \code{extinguish} command and those obtained by propagating the \code{table ISM} field through the uniform slab before reaching the Galactic mid plane.
The main difference between either method and the unextinguished case represented on the top panel of Figure~\ref{figN} is the absence of the inner layer of ionized gas.
Central temperatures and pressures are almost two orders of magnitude lower, and, since there is no phase transition associated to any ionization front, the density of the cold neutral material is equal to $n_0$, also orders of magnitude lower than in the model without attenuation.

The absorption of the ultraviolet photodissociating radiation turns out to be very important as well.
The location of the transition from the cold to the warm neutral phase may change by a large factor, but the most relevant issue is the predicted fraction of Hydrogen in molecular form.
In particular, the \code{extinguish} command overestimates the photodissociation rate so much that there is virtually no molecular Hydrogen at all.
Our results can only strengthen the advice given in Hazy of modelling the absorbing material rather than using the \code{extinguish} command.

\subsection{Stellar mass and Galactic magnetic field}

\begin{figure}
\centerline{ \includegraphics[width=0.5\textwidth]{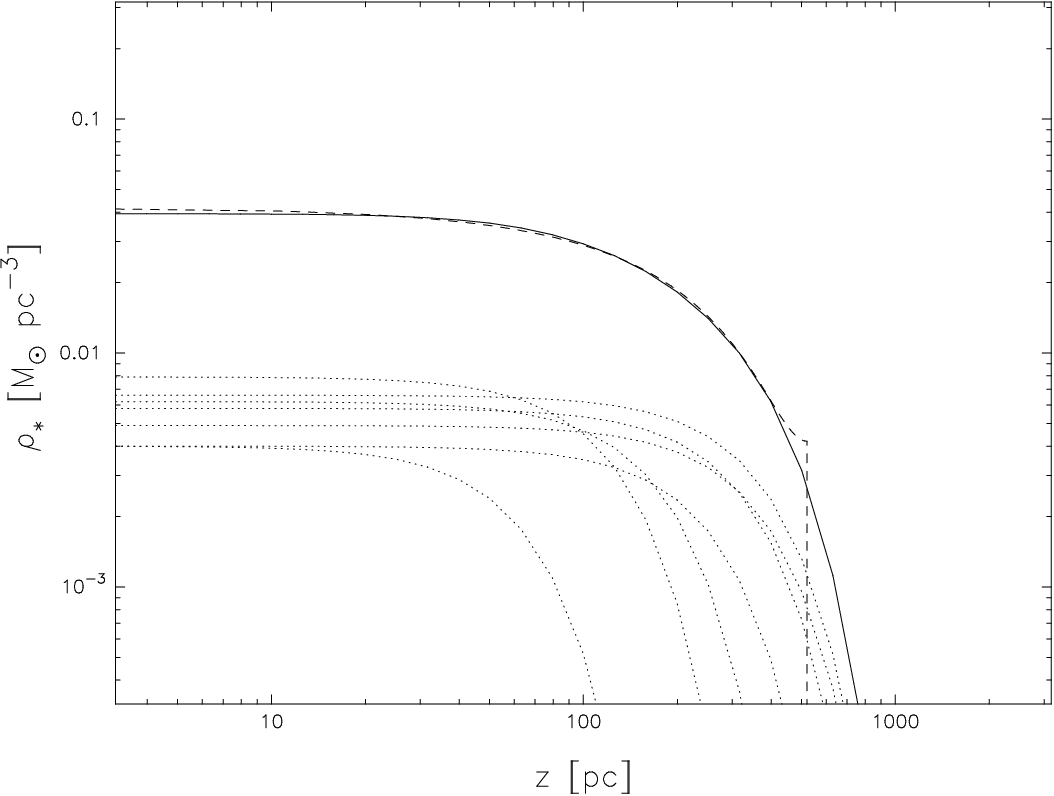} }
\caption
{
Total stellar mass density (solid line) and contributions of individual populations with different ages (dotted lines) according to \citet{Robin+03}.
The approximation given in equation~(\ref{eqRhoStar}) is shown as a dashed line.
}
\label{figRhoStar}
\end{figure}

\begin{figure}
\centerline{ \includegraphics[width=0.5\textwidth]{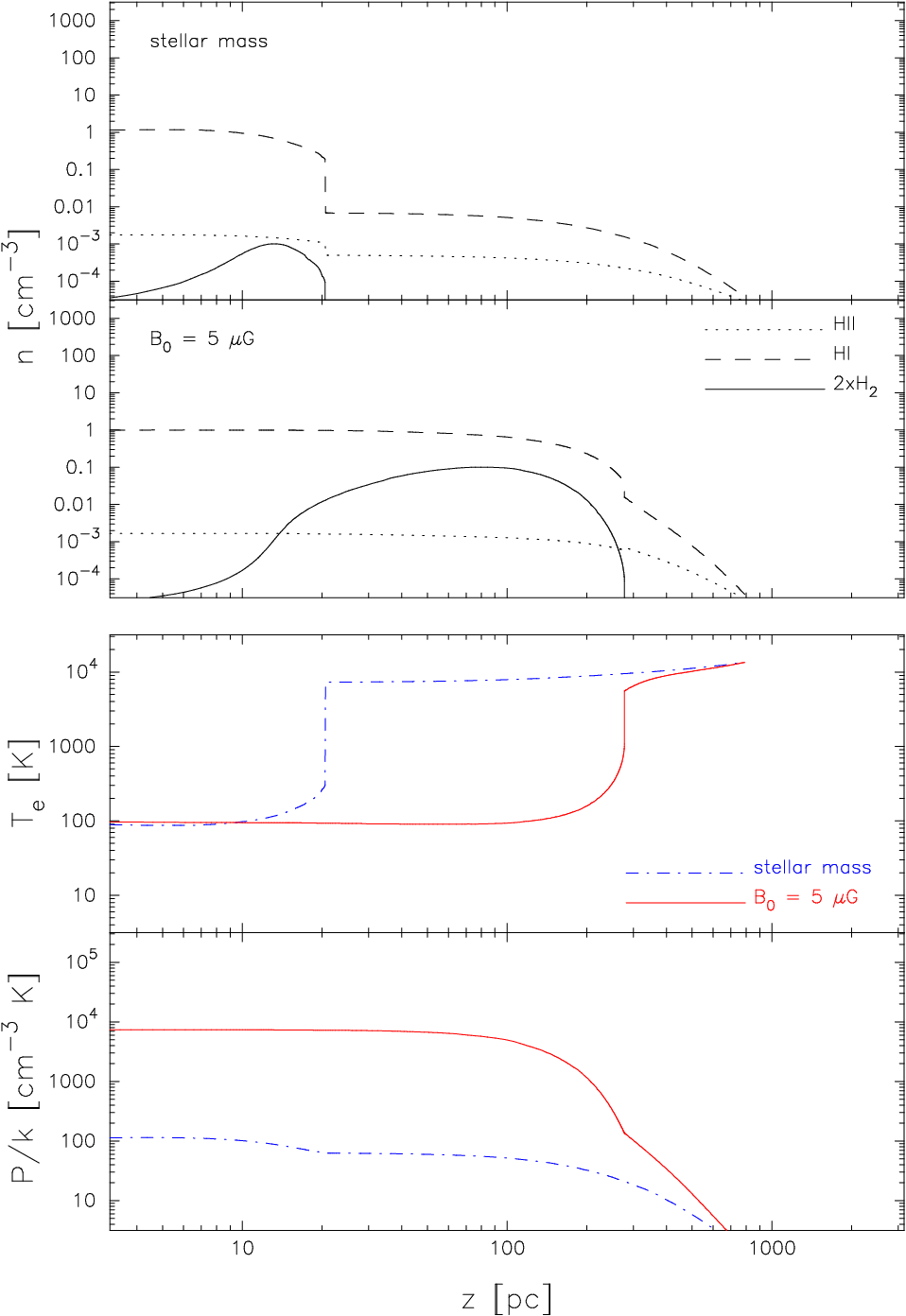} }
\caption
{
Densities of the different Hydrogen phases (top panels), temperatures (third panel) and pressures (bottom panel) for two models including a stellar mass component of the form~(\ref{eqRhoStar}).
In addition, one of them also includes a tangled magnetic field with an intensity $B_0=5~\mu$G at the Galactic plane.
}
\label{figRhoStar_B}
\end{figure}

We have shown that photoionized regions must have a finite extent because of self-gravity.
The exact value of the characteristic scale length depends, however, on many other factors.
For instance, the gas may represent only a small fraction of the total mass of the system, and there can be a significant amount of non-thermal pressure support from turbulent motions, cosmic rays, and magnetic fields \citep{AbelFerland06,Pellegrini+07,Pellegrini+09}.

In the case of the Milky Way, the distribution of stellar mass in the Galactic disk can be inferred by combining number counts of individual stars and dynamical arguments.
We approximate the results obtained by \citet{Robin+03} with the expression
\begin{equation}
 \rho_*(z) = \rho_0\ \left[ 1 - 0.9 \left( \frac{2z}{h_*} - \frac{z^2}{h_*^2} \right) \right]
\label{eqRhoStar}
\end{equation}
where $\rho_0 = 0.042$~M$_\odot$~pc$^{-3}$ and $h_*=520$~pc.
This function, plotted as a dashed line in Figure~\ref{figRhoStar}, provides a reasonable approximation for $z<h_*$, well beyond the half-width half-maximum of the distribution\footnote{In fact, $\rho_*(h_*)=0.1\rho_0$}.
At larger distances, we take $\rho_* \approx 0$.
The adopted value of the central stellar density $\rho_0$ implies the same mass as $1.7$~cm$^{-3}$ of pure Hydrogen, or $\sim1.2$~cm$^{-3}$ in the case of the ISM, where Hydrogen accounts for $\sim72$~per cent of the gas mass.
The contribution of stars to the local density near the Sun is thus comparable to that of the gaseous component.
The cumulative surface density
\begin{equation}
 \Sigma_*(z) = 2\rho_0\ \left[ z - 0.9 \left( \frac{z^2}{h_*} - \frac{z^3}{3h_*^2} \right) \right]
\label{eqSigmaStar}
\end{equation}
increases up to $\sim 17.5$~M$_\odot$~pc$^{-2}$ at $z=h_*$, about 3 times the observed surface density of Hydrogen (in all forms) at the solar radius, and 2 times that of the ISM gas, adding up the mass in Helium and heavier elements.

When the contribution of the stellar component to the gravitational acceleration is taken into account, the pressure gradient required to maintain hydrostatic equilibrium increases, and the transition between the cold and the hot phase moves inwards by about a factor of 2 (top panel in Figure~\ref{figRhoStar_B}).
In general terms, the presence of an external gravitational potential will always push the gas distribution inwards, leading to a reduction of the relevant scale lengths and the total gas mass obtained for the system.

On the other hand, magnetic fields tend to counteract the action of gravity.
Due to the effect of line dragging, the intensity of a tangled field increases with gas density as $B \propto n^{2/3}$ \citep[observationally, $B \propto n^\kappa$, with $0.5<\kappa<1$, see e.g.][]{Crutcher99,Henney+05}, and the associated magnetic pressure varies as $P_{\rm mag} \propto B^2 \propto n^{4/3}$.
This equation of state is stiffer than that of an ideal gas, and therefore a much milder change in gas density is necessary in order to achieve hydrostatic equilibrium.

The second panel of Figure~\ref{figRhoStar_B} shows how the inclusion of a tangled magnetic field with mid-plane intensity $B_0=5~\mu$G \citep{Beck01,HeilesCrutcher05} changes the structure of our ISM model.
As can be inferred from the bottom panels, the magnetic field provides most of the pressure support against gravity.
The extent of the cold phase increases by roughly an order of magnitude, reaching now hundreds of pc.
In the outer, hot layer, pressure is dominated by the thermal component, and the magnetic field (for our adopted value of $B_0$) does not play a significant role.

\subsection{Comparison with observations}

\begin{figure}
\centerline{ \includegraphics[width=0.5\textwidth]{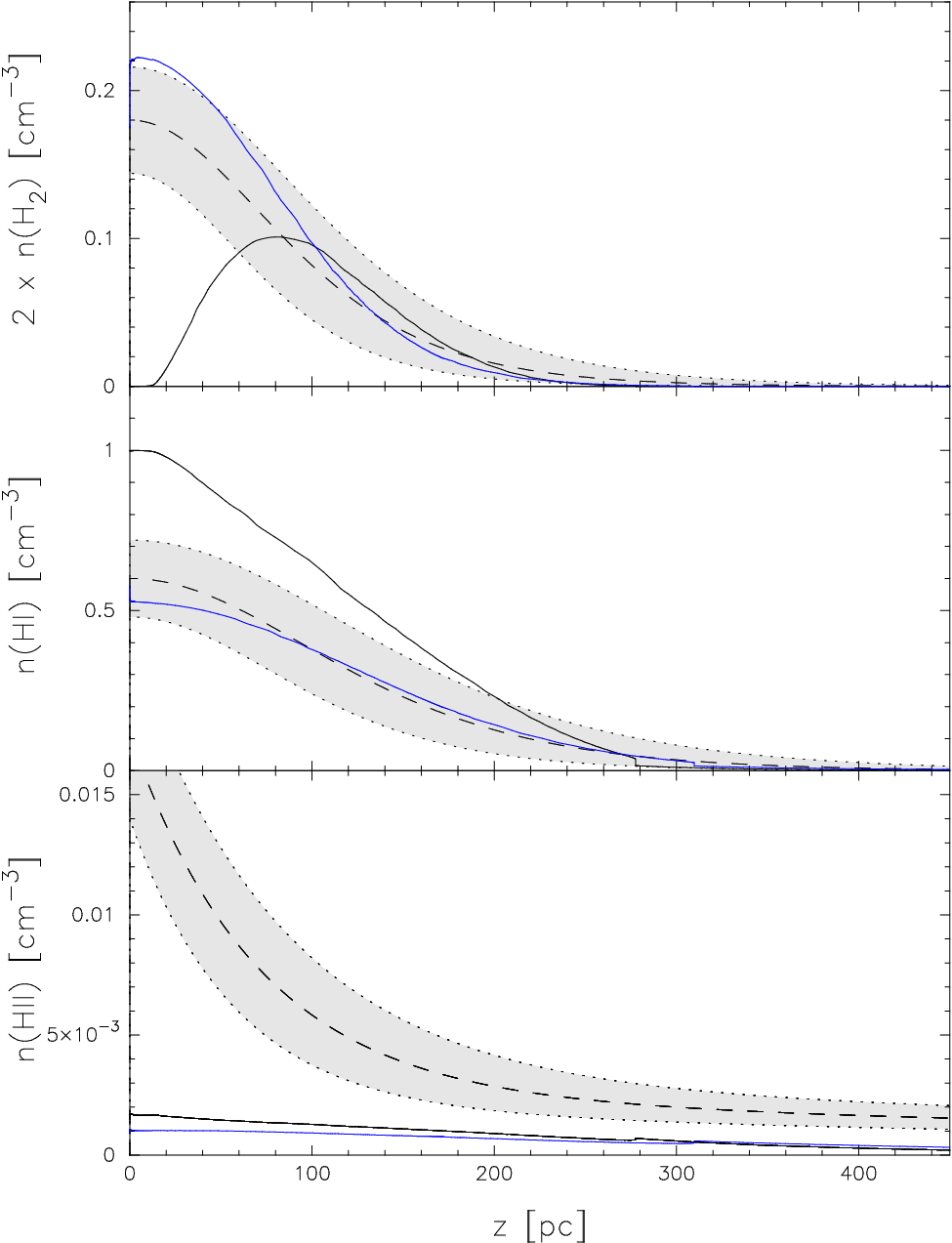} }
\caption
{
Comparison between the model predictions and the observed densities of molecular (top), atomic (middle) and ionized (bottom) Hydrogen.
The observational data are represented as dashed lines, with the grey areas delimited by dotted lines corresponding to 20 per cent error bands.
Black solid lines display the results of the model with $n_0=1$~cm$^{-3}$, $B_0=5~\mu$G, and stellar density~(\ref{eqRhoStar}).
A model with $n_0=0.75$~cm$^{-3}$, $B_0=4~\mu$G, $\kappa=0.5$, and $N({\rm H})=10^{22}$~cm$^{-2}$ is plotted in solid blue (see text for details).
}
\label{figObs}
\end{figure}

Our single-phase model can only describe the average properties of the ISM.
More precisely, it should be able to predict the density of molecular, atomic, and ionized Hydrogen as a function of the distance $z$ from the mid plane of the Galaxy, averaged over some scale larger than the typical size of the inhomogeneities.

Much observational work has been devoted to the determination of the average vertical structure of the Milky Way ISM near the solar neighbourhood.
For the ionized component, we will adopt the distribution derived by \citet{Reynolds91}
\begin{equation}
\frac{ n_{\rm HII}(z) }{\rm cm^{-3}}
= 0.015 \,\exp\! \left(\! -\frac{z}{\rm 70~pc}  \right)
+ 0.0025\,\exp\! \left(\! -\frac{z}{\rm 900~pc} \right)
\end{equation}
from pulsar dispersion measures, where the first term represents the contribution from discrete {\sc Hii} regions near the Galactic plane, and the second term corresponds to the diffuse ionized gas.
For the neutral Hydrogen, we use the formula
\begin{equation}
n_{\rm HI, H_2}(z) = n_0 \,{\rm sech}^2 \left[ \frac{z}{z_{1/2}}\ln(1+\sqrt{2}) \right]
\end{equation}
with $n_0=0.6$~cm$^{-3}$ and $z_{1/2}=125$~pc for the atomic component \citep{NakanishiSofue03}, whereas $n_0=0.09$~cm$^{-3}$ and $z_{1/2}=93$~pc describe the molecular gas \citep{NakanishiSofue06}.
The corresponding density distributions are shown in Figure~\ref{figObs} by dashed lines, and the effect of 20 per cent errors in each normalization and scale length is represented by shaded areas enclosed by dotted lines.
This estimate roughly corresponds to the actual errors quoted in \citet{Reynolds91}, and it is substantially smaller than those in \citet{NakanishiSofue06}.
No information about the observational errors is provided in \citet{NakanishiSofue03}, but, given the large dispersion in the Hydrogen column densities along different lines of sight \citep{DickeyLockman90}, as well as the amplitude of the fluctuations in the azimuthally-averaged surface density profiles, 20 per cent is arguably a realistic estimate.

The results of a model with central density $n_0=1$~cm$^{-3}$, central magnetic field $B_0=5~\mu$G, and a stellar distribution given by equation~(\ref{eqRhoStar}) are also shown in Figure~\ref{figObs} as black solid lines.
Although the model gives a reasonable order-of-magnitude estimate (in agreement with the observed values within a factor of two) of the scale height of the Galactic disk, as well as the densities and total masses of both atomic and molecular Hydrogen, there are some important discrepancies: first, the molecular fraction near the mid plane is severely underestimated; second, the central density is almost a factor of 2 above the observed one; and third, the amount of diffuse ionized gas is extremely low over the whole radial range, especially in the inner regions.

The first two problems are closely related.
One could easily specify a lower mid plane density, but then the amount of molecular Hydrogen would decrease to unacceptably low levels.
It seems that, even after transmission through a column density of $10^{21}$~cm$^{-2}$, the intensity of the photodissociating continuum is still too high for molecules to form at the observed rate.
The third problem is also related to the simple prescription adopted for the incident continuum, since almost all the ionizing radiation becomes absorbed before reaching the illuminated face.
In reality, young stars are spread around the Galactic plane.
Adding any thin source of ionizing photons at the mid plane results in an {\sc Hii} region qualitatively similar to the one observed in Figure~\ref{figN}; in order to reproduce the observed ionization fraction, one would need to specify a distributed radiation source, but such feature is unfortunately not currently implemented in {\sc Cloudy}.

The values of the model parameters can be tuned to better match the observed total density profiles and the molecular Hydrogen fraction.
Blue solid lines in Figure~\ref{figObs} shows the results of a model where the mid-plane density is set to $n_0=0.75$~cm$^{-3}$, the intensity of the magnetic field is $B_0=4~\mu$G at the centre and varies with gas density as $(B/B_0) = (n/n_0)^{0.5}$, and the incident continuum has been propagated through a column density of $N({\rm H})=10^{22}$~cm$^{-2}$.
Nevertheless, we strongly advise caution when interpreting these results.
Although the fair agreement obtained for the atomic and molecular data provides encouraging support for the validity of our implementation of gravity, a more realistic model of the ISM is clearly required in order to derive robust quantitative constraints on its physical properties.
Based on our crude model, we would simply argue that reasonable values of the input parameters (in particular, central gas density and magnetic field) yield reasonable scale heights for the disk.
Atomic and molecular Hydrogen densities typically agree with observations within a factor of two, but a prescription for distributed sources is necessary in order to model the ionized component.

\section{Conclusions}
\label{secConclusions}

The present \Revised{work} describes an implementation of gravity in the well-tested photoionization code {\sc Cloudy}.
For an atmosphere in hydrostatic equilibrium, the gravitational force has to be balanced by a pressure gradient according to equation~(\ref{eqHydro}).

By way of example, we consider a plane-parallel model of the Galactic ISM in the solar neighbourhood to illustrate the influence of the gravitational acceleration on the physical properties of the gas.
The main result, valid in the general case, is that gravity sets the scale height of the system.
If self-gravity is not accounted for, there is nothing that prevents the gas (mostly molecular at large radii) from extending to infinity.
When gravity is present, the pressure has to decrease in order to maintain hydrostatic equilibrium, and the gas becomes less dense.
The precise structure depends on many other variables; we have investigated the effect of \Revised{the incident radiation field, the presence of an external mass, and} a tangled magnetic field, but other ingredients, such as turbulent motions, cosmic rays, dust grains, or variations in the atomic rate coefficients \citep[see e.g.][]{Abel+08}, as well as the external pressure of the intergalactic medium \citep{SilichTenorio-Tagle01} may also play an important role.
\Revised
{
To summarize, the present results show that gravity (in particular, the self-gravity of the gas) should not be neglected in studies of the structure of the atomic and molecular layers of a photoionized region, especially when one is interested in their total extent.

Concerning the specific problem of the vertical structure of the Milky Way ISM, the models discussed here are able to reproduce the observed distribution of atomic and molecular Hydrogen with an accuracy of about a factor of two, but the amount of diffuse ionized gas is severely underestimated at all heights.
Much more realistic results could be obtained by providing support for distributed radiation sources, as well as a treatment for the coexistence of different gas phases at the same point.
}

\section*{Acknowledgments}

We would like to thank G.~Ferland and the {\sc Cloudy} team for developing, maintaining, and making the software publicly available.
It is also a pleasure to acknowledge useful comments from G.~Tenorio-Tagle and E.~P\'erez on a preliminary version of this manuscript.
Implementation in the development version of {\sc Cloudy} would not be possible without the help from G.~Ferland, R.~Williams, R.~Porter, and P.~van~Hoof.
\Revised{We also thank the referee, W. Henney, for his insightful comments and suggestions, leading to major improvements in our Milky Way model.}
Financial support for this work has been provided by the Spanish \emph{Ministerio de Educaci\'on y Ciencia} through project AYA2007-67965-C03-03.

 \bibliographystyle{mn2e}
 \bibliography{references}

\begin{thebibliography}{}

\bibitem[\protect\citeauthoryear{{Abel} \& {Ferland}}{{Abel} \&
  {Ferland}}{2006}]{AbelFerland06}
{Abel} N.~P.,  {Ferland} G.~J.,  2006, \apj, 647, 367

\bibitem[\protect\citeauthoryear{{Abel}, {Hoof}, {Shaw}, {Ferland} \&
  {Elwert}}{{Abel} et~al.}{2008}]{Abel+08}
{Abel} N.~P.,  {Hoof} P.~A.~M.~v.,  {Shaw} G.,  {Ferland} G.~J.,    {Elwert}
  T.,  2008, \apj, 686, 1125

\bibitem[\protect\citeauthoryear{{Beck}}{{Beck}}{2001}]{Beck01}
{Beck} R.,  2001, Space Science Reviews, 99, 243

\bibitem[\protect\citeauthoryear{{Black}}{{Black}}{1987}]{Black87}
{Black} J.~H.,  1987, in {Hollenbach} D.~J.,  {Thronson} Jr. H.~A.,  eds,
  Interstellar Processes Vol.~134 of Astrophysics and Space Science Library,
  {Heating and cooling of the interstellar gas}.
pp 731--744

\bibitem[\protect\citeauthoryear{{Boulares} \& {Cox}}{{Boulares} \&
  {Cox}}{1990}]{BoularesCox90}
{Boulares} A.,  {Cox} D.~P.,  1990, \apj, 365, 544

\bibitem[\protect\citeauthoryear{{Cordes} \& {Lazio}}{{Cordes} \&
  {Lazio}}{2002}]{CordesLazio_02}
{Cordes} J.~M.,  {Lazio} T.~J.~W.,  2002, arXiv:astro-ph/0207156

\bibitem[\protect\citeauthoryear{{Cowie} \& {Songaila}}{{Cowie} \&
  {Songaila}}{1986}]{CowieSongaila86}
{Cowie} L.~L.,  {Songaila} A.,  1986, \araa, 24, 499

\bibitem[\protect\citeauthoryear{{Cox}}{{Cox}}{2005}]{Cox05}
{Cox} D.~P.,  2005, \araa, 43, 337

\bibitem[\protect\citeauthoryear{{Crutcher}}{{Crutcher}}{1999}]{Crutcher99}
{Crutcher} R.~M.,  1999, \apj, 520, 706

\bibitem[\protect\citeauthoryear{{Dickey} \& {Lockman}}{{Dickey} \&
  {Lockman}}{1990}]{DickeyLockman90}
{Dickey} J.~M.,  {Lockman} F.~J.,  1990, \araa, 28, 215

\bibitem[\protect\citeauthoryear{{Elitzur} \& {Ferland}}{{Elitzur} \&
  {Ferland}}{1986}]{ElitzurFerland86}
{Elitzur} M.,  {Ferland} G.~J.,  1986, \apj, 305, 35

\bibitem[\protect\citeauthoryear{{Ferland} \& {Elitzur}}{{Ferland} \&
  {Elitzur}}{1984}]{FerlandElitzur84}
{Ferland} G.~J.,  {Elitzur} M.,  1984, \apjl, 285, L11

\bibitem[\protect\citeauthoryear{{Ferland}, {Korista}, {Verner}, {Ferguson},
  {Kingdon} \& {Verner}}{{Ferland} et~al.}{1998}]{Ferland+98}
{Ferland} G.~J.,  {Korista} K.~T.,  {Verner} D.~A.,  {Ferguson} J.~W.,
  {Kingdon} J.~B.,    {Verner} E.~M.,  1998, \pasp, 110, 761

\bibitem[\protect\citeauthoryear{{Ferland} \& {Mushotzky}}{{Ferland} \&
  {Mushotzky}}{1984}]{FerlandMushotzky84}
{Ferland} G.~J.,  {Mushotzky} R.~F.,  1984, \apj, 286, 42

\bibitem[\protect\citeauthoryear{{Ferri{\`e}re}}{{Ferri{\`e}re}}{2001}]{Ferrie%
re01}
{Ferri{\`e}re} K.~M.,  2001, Reviews of Modern Physics, 73, 1031

\bibitem[\protect\citeauthoryear{{Heiles} \& {Crutcher}}{{Heiles} \&
  {Crutcher}}{2005}]{HeilesCrutcher05}
{Heiles} C.,  {Crutcher} R.,  2005, in {R.~Wielebinski \& R.~Beck} ed., Cosmic
  Magnetic Fields Vol.~664 of Lecture Notes in Physics, Berlin Springer Verlag,
  {Magnetic Fields in Diffuse HI and Molecular Clouds}.
pp 137--+

\bibitem[\protect\citeauthoryear{{Henney}, {Arthur}, {Williams} \&
  {Ferland}}{{Henney} et~al.}{2005}]{Henney+05}
{Henney} W.~J.,  {Arthur} S.~J.,  {Williams} R.~J.~R.,    {Ferland} G.~J.,
  2005, \apj, 621, 328

\bibitem[\protect\citeauthoryear{{Mathis}, {Rumpl} \& {Nordsieck}}{{Mathis}
  et~al.}{1977}]{Mathis+77}
{Mathis} J.~S.,  {Rumpl} W.,    {Nordsieck} K.~H.,  1977, \apj, 217, 425

\bibitem[\protect\citeauthoryear{{Meyer}, {Jura} \& {Cardelli}}{{Meyer}
  et~al.}{1998}]{Meyer+98}
{Meyer} D.~M.,  {Jura} M.,    {Cardelli} J.~A.,  1998, \apj, 493, 222

\bibitem[\protect\citeauthoryear{{Miller} \& {Cox}}{{Miller} \&
  {Cox}}{1993}]{MillerCox93}
{Miller} W.~W.~I.,  {Cox} D.~P.,  1993, \apj, 417, 579

\bibitem[\protect\citeauthoryear{{Nakanishi} \& {Sofue}}{{Nakanishi} \&
  {Sofue}}{2003}]{NakanishiSofue03}
{Nakanishi} H.,  {Sofue} Y.,  2003, \pasj, 55, 191

\bibitem[\protect\citeauthoryear{{Nakanishi} \& {Sofue}}{{Nakanishi} \&
  {Sofue}}{2006}]{NakanishiSofue06}
{Nakanishi} H.,  {Sofue} Y.,  2006, \pasj, 58, 847

\bibitem[\protect\citeauthoryear{{Paladini}, {De Zotti}, {Davies} \&
  {Giard}}{{Paladini} et~al.}{2005}]{Paladini+05}
{Paladini} R.,  {De Zotti} G.,  {Davies} R.~D.,    {Giard} M.,  2005, \mnras,
  360, 1545

\bibitem[\protect\citeauthoryear{{Pellegrini}, {Baldwin}, {Brogan}, {Hanson},
  {Abel}, {Ferland}, {Nemala}, {Shaw} \& {Troland}}{{Pellegrini}
  et~al.}{2007}]{Pellegrini+07}
{Pellegrini} E.~W.,  {Baldwin} J.~A.,  {Brogan} C.~L.,  {Hanson} M.~M.,  {Abel}
  N.~P.,  {Ferland} G.~J.,  {Nemala} H.~B.,  {Shaw} G.,    {Troland} T.~H.,
  2007, \apj, 658, 1119

\bibitem[\protect\citeauthoryear{{Pellegrini}, {Baldwin}, {Ferland}, {Shaw} \&
  {Heathcote}}{{Pellegrini} et~al.}{2009}]{Pellegrini+09}
{Pellegrini} E.~W.,  {Baldwin} J.~A.,  {Ferland} G.~J.,  {Shaw} G.,
  {Heathcote} S.,  2009, \apj, 693, 285

\bibitem[\protect\citeauthoryear{{Reynolds}}{{Reynolds}}{1991}]{Reynolds91}
{Reynolds} R.~J.,  1991, in {H.~Bloemen} ed., The Interstellar Disk-Halo
  Connection in Galaxies Vol.~144 of IAU Symposium, {Ionized disk/halo gas -
  Insight from optical emission lines and pulsar dispersion measures}.
pp 67--76

\bibitem[\protect\citeauthoryear{{Robin}, {Reyl{\'e}}, {Derri{\`e}re} \&
  {Picaud}}{{Robin} et~al.}{2003}]{Robin+03}
{Robin} A.~C.,  {Reyl{\'e}} C.,  {Derri{\`e}re} S.,    {Picaud} S.,  2003,
  \aap, 409, 523

\bibitem[\protect\citeauthoryear{{Savage} \& {Sembach}}{{Savage} \&
  {Sembach}}{1996}]{SavageSembach96}
{Savage} B.~D.,  {Sembach} K.~R.,  1996, \araa, 34, 279

\bibitem[\protect\citeauthoryear{{Silich} \& {Tenorio-Tagle}}{{Silich} \&
  {Tenorio-Tagle}}{2001}]{SilichTenorio-Tagle01}
{Silich} S.,  {Tenorio-Tagle} G.,  2001, \apj, 552, 91

\bibitem[\protect\citeauthoryear{{Silich} \& {Tenorio-Tagle}}{{Silich} \&
  {Tenorio-Tagle}}{1998}]{SilichTenorio-Tagle98}
{Silich} S.~A.,  {Tenorio-Tagle} G.,  1998, \mnras, 299, 249

\bibitem[\protect\citeauthoryear{{Taylor} \& {Cordes}}{{Taylor} \&
  {Cordes}}{1993}]{TaylorCordes93}
{Taylor} J.~H.,  {Cordes} J.~M.,  1993, \apj, 411, 674

\end{thebibliography}

\appendix
\section{User Guide}
\label{secUserGuide}

In order to specify a hydrostatic equilibrium configuration, the \code{constant pressure} command must be included in the {\sc Cloudy} input script.
In addition, a \code{gravity} command with syntax
\begin{verbatim}
 gravity <symmetry> [<factor> [_log]]
\end{verbatim}
has been implemented.
The compulsory argument \code{symmetry} must be either \emph{spherical} or \emph{plane-parallel} in order to tell the code whether to use equation~(\ref{eqGrav_sphere}) or (\ref{eqGrav_plane}) to compute the gravitational acceleration\footnote{Actually, the command parser looks for the character sequences ``sphe'' and ``plan'', so expressions like ``sphere'' or ``planar'' are also valid.}.
An optional factor can be introduced that multiplies the gas mass, mimicking the presence of an external component with identical distribution.
A more flexible way to specify an external potential is provided by the command
\begin{verbatim}
 gravity external <mass> [<extent> <index>] [_log]
\end{verbatim}
that models a distribution of mass around the centre of symmetry of the system.
In the spherical case, the first number represents the mass (in solar masses) of a pointlike object located at $r=0$; in the plane-parallel case, it corresponds to the total surface density (in~M$_\odot$~pc$^{-2}$) of a uniform thin sheet at the mid plane $z=0$.
In both cases, the total \code{mass} $m_0$ is assumed to be distributed over an \code{extent} $x_0$ as a power law with \code{index} $\alpha$, i.e. $m(x) = m_0 (x/x_0)^\alpha$.
All numbers are interpreted as linear, unless the keyword \code{\_log} is specified.

The following is an example of a typical input script used in the preparation of this work:
\begin{verbatim}
# -----------------------------------------------
#  Cloudy input file for hydrostatic equilibrium
#                Yago Ascasibar (UAM, Fall 2009)
# -----------------------------------------------
#
c -----------------------------------------------
title - Hydrostatic equilibrium -
c -----------------------------------------------
c Incident continuum
c -----------------------------------------------
# CMB
# table ism
# extinguish column = 21 leak = 0
table read "attenuated_ism.dat"
nuF(nu) = -2.0453234 at 7.541e-05 Ryd
c -----------------------------------------------
c Density and chemical composition
c -----------------------------------------------
hden 1 linear
abundances ism
c -----------------------------------------------
c Gravity
c -----------------------------------------------
constant pressure
gravity plane-parallel
gravity external Sigma = 43.680, z0 = 520, a = 1
gravity external Sigma =-39.312, z0 = 520, a = 2
gravity external Sigma = 13.104, z0 = 520, a = 3
c -----------------------------------------------
c Non-thermal components
c -----------------------------------------------
cosmic ray background
magnetic field -5.3
c -----------------------------------------------
c Stopping conditions
c -----------------------------------------------
stop temperature off
stop thickness 1e4 linear parsec
stop eden -4.5
iterate to convergence
c -----------------------------------------------
c Output files
c -----------------------------------------------
punch overview "overview.txt" last
punch pressure "pressure.txt" last
# punch transmitted continuum "trans.txt" last
# punch hydrogen conditions "Hydrogen.txt" last
c -----------------------------------------------
c                         ... Paranoy@ Rulz! ;^D
c -----------------------------------------------
\end{verbatim}

The first blocks set a label to identify the run, the incident radiation field, the central Hydrogen density $n_0$ and the gas chemical composition.
Commands related to hydrostatic equilibrium come next.
Note, in particular, how several \code{gravity external} commands have been combined in order to specify the stellar mass distribution given by expression~(\ref{eqSigmaStar}).
The integration continues until one of the stopping conditions is met (the disk thickness reaching 10~kpc or the electron density falling below $\sim3\times10^{-5}$~cm$^{-3}$), and it is iterated until the optical depths converge.
Finally, the main physical properties of the gas (e.g. densities, temperatures) as a function of distance, as well as the different contributions to the total pressure, are printed to output files on the disk.

\end{document}